# Monitoring charge separation of individual cells in perovskite/silicon tandems via transient surface photovoltage spectroscopy

Maxim Simmonds [1], Ke Xu [1], Steve Albrecht [1,2], Lars Korte*[1,2], Igal Levine*[3]

[1] Solar Energy Division, Helmholtz-Zentrum Berlin für Materialien und Energie GmbH, Kekuléstraße 5, 12489 Berlin, Germany
[2] Faculty of Electrical Engineering and Computer Science, Technical University Berlin, Marchstraße 23, 10587 Berlin, Germany
[3] Institute of Chemistry and The Center for Nanoscience and Nanotechnology, The Hebrew University of Jerusalem, Jerusalem 9190401, Israel

*Email: igal.levine@mail.huji.ac.il, korte@helmholtz-berlin.de

## Abstract

Identification of charge carrier separation processes in perovskite/silicon tandem solar cells and recombination at buried interfaces of charge selective contacts is crucial for photovoltaic research. Here, intensity- and wavelength-dependent transient surface photovoltage (tr-SPV) is used to investigate slot-die-coated perovskite top layers deposited on n-type Heterojunction Silicon bottom cells. We show that using an appropriate combination of photon energy and/or bottom cell polarity, one can individually probe the buried interfaces of the bottom silicon cell or the perovskite's buried interfaces of a tandem solar cell: For excitation with higher energy photons, time delays before the onset of a strong SPV signal indicate significant hole minority drift before separation in the silicon bottom cells. Furthermore, symmetric bottom Si heterojunction solar cell stacks can serve to investigate the top perovskite stack including its junction to the bottom cell, unhampered by photovoltages from the silicon substrate. Thus, investigation of the buried interfaces in tandem devices using time-resolved surface photovoltage is found to yield valuable information on charge carrier extraction at buried interfaces and demonstrates its unique potential compared to more conventional approaches that rely on photoluminescence decay kinetics.

## 1. Introduction

Perovskite/silicon (Si) tandem solar cells have demonstrated high solar energy conversion efficiencies in the past few years. Most of the highest efficiency solar cells are now based on n-type silicon, marking a significant shift from the previous dominance of p-type wafers in 2024.[1–3] Nowadays, the most efficient solar cells are silicon heterojunction (SHJ) cells using crystalline silicon (c-Si) wafers as the absorber and hydrogenated amorphous or nanocrystalline silicon/silicon oxides (a-Si:H or nc-Si:H/nc-SiOx:H) to form the contacts for charge extraction. Such SHJ cells are also used in today's most efficient perovskite/silicon tandem cells [4], which makes them strong candidates for upscaling and industrialization.[5]

A full tandem solar cell stack is comprised of many different layers that result in numerous interfaces. Optoelectronic characterization of the individual electronic junctions that are formed in each of these interfaces is a challenging task, not only because most of them are buried interfaces, but also because they can be affected by the overall built-in field within the device. Using Kelvin probe and photoelectron yield spectroscopy to measure SPV, we recently demonstrated the importance of the built-in field (or: voltage) in the perovskite cell device stack for charge carrier separation and its crucial dependence on the contact metal's work function.[6] To date, most optoelectronic characterization of tandem solar cells is done by absolute or transient photoluminescence (tr-PL)[7–9] and current (density)-voltage (J-V) measurements both in the dark and under illumination. These methods enable investigation of various properties but suffer several drawbacks. To name a few, J-V characterization requires fabrication of the entire device including electrical contacts, which does not allow investigation of a single interface. Using tr-PL, one cannot disentangle interfacial defect-assisted recombination from charge extraction (e.g. blocked transport by energetic misalignment at the interfaces) without additional measurements.[10] Thus, understanding fill factor losses and quantifying the charge carrier dynamics at a specific interface over various time regimes of full tandem solar cells remains a longstanding challenge.

This work is dedicated to analyzing the charge carrier dynamics occurring in slot-die coated perovskite thin film coated on two different bottom Si heterojunction structures: (1) a Heterojunction solar cell on n-type c-Si, and (2) a c-Si (n)/ nc-SiOx:H (n+) symmetric Si Heterojunction, by utilizing excitation wavelength and intensity-dependent, transient Surface PhotoVoltage (tr-SPV) measurements. In contrast to J-V and PL measurements, SPV measurements do not require electrical contacts and do not rely on radiative recombination. An SPV signal is generated upon light-induced separation of charge carriers, with the direction of charge separation given by the sign of the SPV signal.[11] A positive (negative) SPV signal indicates the separation of holes (electrons) being displaced toward the SPV probe's surface. Our previous study on single-junction perovskite solar cells demonstrated that tr-SPV measurements yield valuable information on the charge carrier dynamics in buried interfaces over a wide time range from nanoseconds to seconds.[10] Varying the excitation wavelength allows exciting separately the Si bottom cell or the perovskite top cell, provided that the top cell absorbs close to all photons excited. Such selective excitation of the bottom and top cells individually was recently applied using Electrochemical Impedance Spectroscopy by Roose et al.,[11] demonstrating how the response of the individual cells can be recorded separately, yet it still requires an often-complex electrical circuit modelling, which can be omitted by using tr-SPV.

First, we show an energy band diagram for the Si bottom cells to present similarities and differences in the different stacks used. Next, wavelength-dependent and intensity-dependent tr-SPV are recorded and investigated for the different layer stacks. Last, based on the differences in the tr-SPV transients between the samples, a model that describes the charge separation and recombination processes is introduced and discussed.

## 2. Results and Discussion: Tandem stacks and band alignment

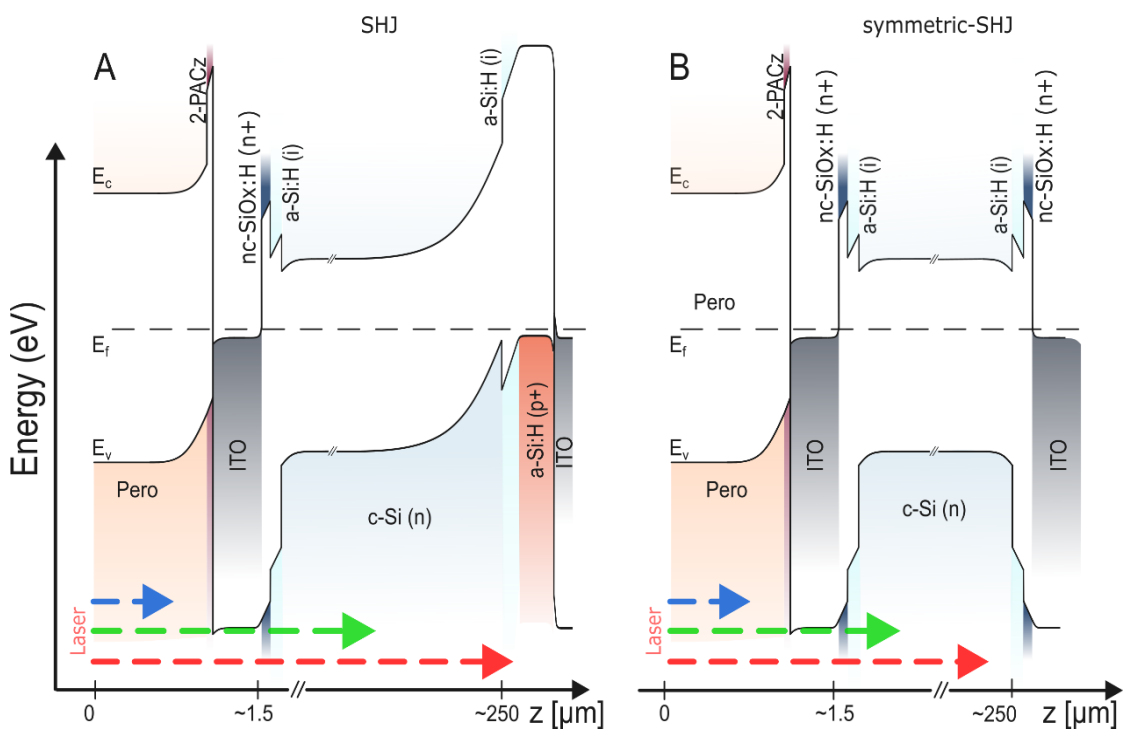

Figure 1: Sketch of the estimated energy band diagrams (neglecting interface dipole for the 2-PACz layer), including the location of the SPV probe and illumination direction; (A) SHJ: n-type Floatzone (FZ) Si (c-Si(n)) Heterojunction with an a-Si:H (p+) layer on the rear side of the Si bottom cell (SHJ) and a slot-die coated 3haldie perovskite top cell. (B) symmetric-SHJ: n-type FZ Si (c-Si(n)) Heterojunction cell with nc-SiOx:H (n+) layers on both sides of the Si wafer and a slot-die coated 3halide perovskite top cell. In both cases, the Si wafer was ~250 µm thick.

The two types of tandem architectures are shown in Figure 1 (A) and (B) for the SHJ and symmetric-SHJ, respectively. In both cases, the light absorber of the top cells is an 800 nm thick 3halide perovskite layer ($Cs_{0.22}FA_{0.78}Pb(I_{0.85}Br_{0.15})_3$ + 5 mol% $MAPbCl_3$ (3halide perovskite) with 1.68 eV band gap. As a hole transport layer (HTL) to the perovskite layer, 2PACz ([2-(9H-Carbazol-9-yl)ethyl]phosphonic Acid) was used. The key difference between the SHJ and symmetric-SHJ stems from the presence or absence of the p+/n selective junction (hole contact) at the back of the tandem stack. The nc-SiOx:H(n) is a surface field layer that minimizes surface recombination and improves the ohmic contact to the ITO layer. Analogously to 2PACz in perovskite solar cells as hole selective layer, the nc-SiOx(n) acts as charge selective layer (for electrons, in this case), but provides almost no built-in field. The n+/n/n+ stack in Figure 1(B) is thus an "all-electron" majority carrier device. Electrons can easily flow through it towards the tunnel-recombination junction with the perovskite top cell, but due to its symmetry and negligible band bending at both interfaces, no (pronounced) net SPV signal can build up in the n+/n/n+ stack. The lack of band bending is exemplified in Figure S3 (comparison of p+/n/n+ versus n+/n/n+), where the SPV amplitude is greatly reduced for the symmetric-SHJ solar cell.

Next, we focus our attention on tr-SPV measurements for both the SHJ and the symmetric-SHJ bottom cells, covered with the thick 3halide perovskite layers.

### 2.1 Wavelength-dependent SPV

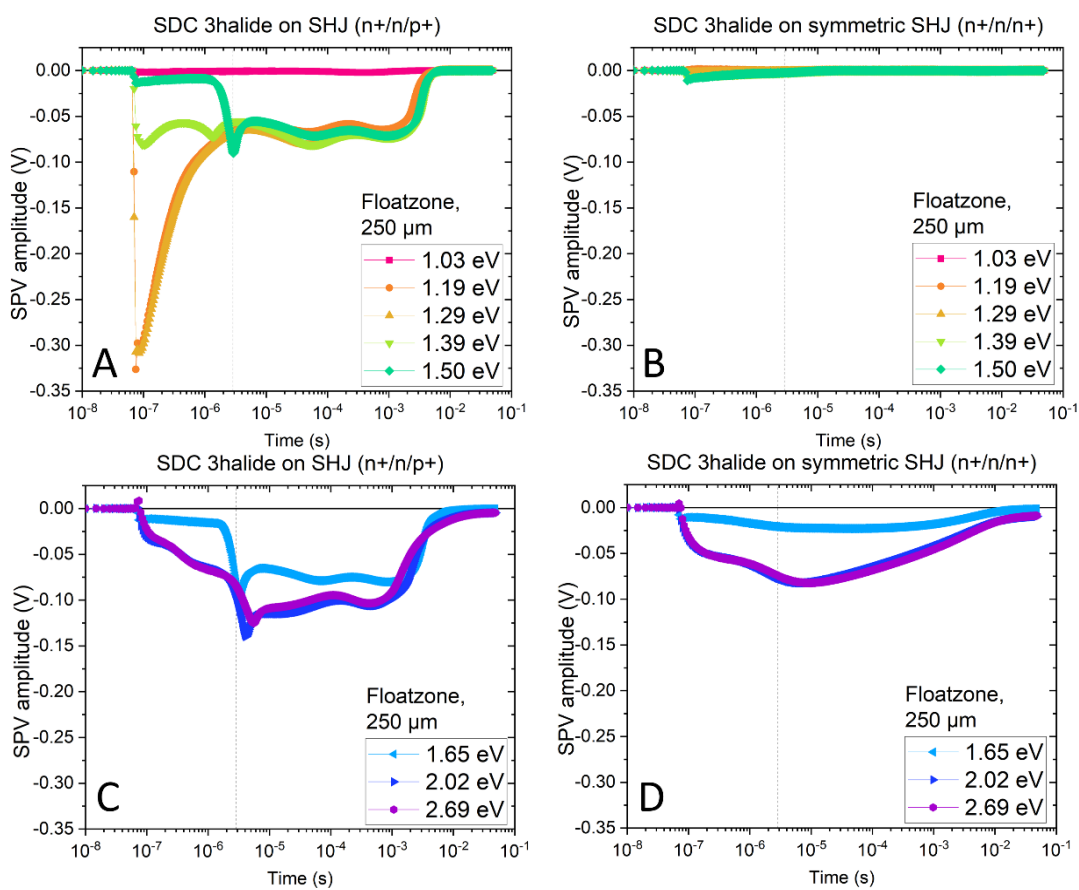

Figure 2: Transient-SPV signals for SHJ (A, C) and symmetric SHJ (B, D) using front (slot-die coated 3halide perovskite side) illumination at different excitation wavelengths (1.03-2.69 eV) and a repetition rate of 2 Hz. The silicon substrate is approximately 250 μm thick.

To study the charge carrier dynamics in the individual sub-cells of the tandem structure, wavelength-dependent tr-SPV measurements were performed as shown in Figure 2. By selecting the laser excitation wavelength, we selectively excite either only the bottom cell, or mainly the top perovskite cell. To obtain a good signal-to-noise ratio across all photon energies, relatively high laser fluences were used. However, we ensured that the photon flux is nearly constant at different wavelengths, as shown in Figure S1 in the Supplementary information. All observed SPV values were negative as expected: since both sub-cells have holes extracted towards their respective rear contact (right side Figures 1A and 1B), while electrons are separated toward the front side where the SPV electrode is located.

Starting with the tandem with an SHJ bottom cell in Figure 2 (A), excitation from 1.03 to 1.50 eV is purely absorbed by the Si absorber (for the EQE spectra of both sub-cells, see Figure S2 in the Supplementary information). At 1.19 eV, the inverse of the absorption coefficient is $\alpha^{-1} \approx 1000$ μm for Si [12], meaning the generation profile in the silicon wafer is homogenous. For this case, we observe a sharp and fast initial SPV peak of 326 mV. A higher photon energy of 1.50 eV (where $\alpha^{-1} \approx 19$ μm, with carriers generated towards the front contact Si junction), results initially in a small negative SPV of around 12 mV, which remains nearly constant for around 3 μs. We explain this small first initial SPV as corresponding to the reduction of the weaker front contact band bending in the Silicon absorber. This leads to extraction of electrons into the front recombination ITO layer. Afterward, a sharp, delayed second rise in the SPV is observed. The SPV signal remains rather constant up to 3 μs. This suggests that the number of photogenerated electrons and holes, which were separated in space, did not increase during this period. This is because in the SHJ, the electron-hole pairs photo-generated at the Si front side undergo ambipolar diffusion into the Si bulk with different electron and hole mobilities until they reach the rear side selective layer (hole contact), reducing the stronger back contact band bending and leading to hole extraction, generating a second delayed and stronger rise in SPV. Under similar photon fluence, the SPV with 1.19 eV excitation attains a peak value of -326 mV whereas the 1.39 eV excitation yields only -70 mV after 2 μs. This can be explained due to recombination of electron-hole pairs during the diffusion of these carriers to the rear n/p+ interface. Note that the recombination of electron-hole pairs during diffusion will correspond to decreased performance of a solar cell. Hence, assuming our interpretation of the results is correct, the tr-SPV measured on the perovskite/SHJ stack at a photon energy of 2.69 eV indicates that the back contact band bending is suppressed due to the arrival and extraction of holes 6 μs after the excitation pulse (SPV maximum in Figure 2C, purple curve). To quantitatively confirm this interpretation, diffusion simulations were conducted (Supplementary Notes SI.6 and SI.7). The differential equations ES1 and ES2 account for carrier diffusion as well as non-radiative Shockley-Read-Hall (defect recombination) and Auger processes. These drift simulations show that the ambipolar arrival time is approximately $\tau_{ambipolar} = 17$ μs (see Figure S6C), meaning that carriers require on average 17 μs to diffuse through the full wafer thickness (250 μm). This value is about three times larger than the experimentally observed SPV peak time, $\tau_{exp} = 6$ μs.

Furthermore, at 6 μs after the laser pulse, carriers have diffused an average distance of

$$L_{exp} = \sqrt{2D_{ambipolar}\tau_{exp}} = 130 \text{ μm}, \quad \text{as derived from Fick's law, using}$$

$D_{ambipolar} = 14.3 \ cm^2 s^{-1}$ (ambipolar diffusion coefficient from [13]). Hence, only carriers photogenerated within 130 µm from the back can be responsible for suppressing the back contact band bending, before the rest of the carriers—generated further toward the front—arrive.

To verify this hypothesis, we evaluate whether there are enough photogenerated carriers within this region to compensate the band bending. For this, we integrate the carrier density over the final 130 µm of the wafer and assume these carriers reside within the back contact band bending region. This yields: $\Delta p_{back} = \frac{\int_{L-L_d}^{L} \Delta p(x) \, dx}{L_{inversion}} = 2.2 \ 10^{16} cm^{-3}$, with $L_{inversion} \approx 500 \ nm$ [14], corresponding to the thickness of the back contact band bending region . This excess carrier concentration corresponds to a quasi-Fermi level splitting (QFLS) calculated as: $QFLS = kT \ln \left( \frac{\Delta p_{back}^2}{n_i^2} \right) \approx 750$ meV, which is in close agreement with an expected value for the built-in potential of an SHJ (~800 meV) [15]. Therefore, there are sufficient initially photogenerated carriers in the tail of the pulsed laser generation profile (x > 120 µm) to cancel a significant portion of the back contact band bending. This supports our hypothesis that photogenerated carriers deeper within the wafer are responsible for suppressing the back contact band bending, thereby explaining the earlier experimental SPV peak ($\tau_{exp} < \tau_{ambipolar}$).

Ambipolar diffusion of charge carriers can also explain the shifts in the observed times of the second SPV peaks of Fig 2A. As the excitation wavelength decreases, photogeneration takes place further away from the rear side, resulting in larger diffusion lengths and hence longer diffusion times (see the two green curves of Fig 2A, as well as Figure S3A). To further prove that the ambipolar diffusion plateau is related to the Si bottom cell only, an independent measurement was conducted on a 250 µm Floatzone Heterojunction Si substrate (without a perovskite layer, shown in Figure S3 A and B), where similar charge carrier dynamics were observed. The diffusion simulation results of Figure S6 also show that the resulting electron arrival times indeed increase as a function of photon energy and will saturate for photon energies larger than 1.5 eV. The saturation value at E > 1.5 eV in Figure S6C is consistent with the saturation of the SPV peak time. Thus, based on **a)** the experimental confirmation that an SHJ-only cell exhibits an SPV plateau at early times (see Figure S3A), along with **b)** simulation results that align with the experimentally observed increase in carrier arrival time as a function of photon energy, and **c)** the observed decrease in arrival time with increasing pulse power (Figure 3, with further discussion in the next section), we conclude that the delayed SPV rise is indeed caused by ambipolar diffusion of photogenerated electrons and holes toward the rear side of the silicon wafer.

For the SHJ tandem, in the 1.6–2.7 eV range (Figure 2C), as the excitation wavelength decreases, more and more electron-hole pairs are generated in the perovskite layer. Therefore, the initial peak at 0.37 µs (which is absent from the Si-only SHJ cell as shown in Figure S3 A and B, as expected) is related to hole extraction to the HTL underneath the perovskite layer. The holes extracted from the perovskite layer then recombine at the interface to the n-type ITO layer, which supplies the electrons required as recombination partners. Note that in the measurement conditions presented, we still observe the delayed SPV peak (at 1μs) linked to the extraction of holes at the n/p+ interface on the rear of the silicon bottom cell.

To further separate the different contributions of the perovskite/HTL hole extraction interface and the silicon n/p+ interface, tr-SPV was also measured for symmetric-SHJ stacks as substrates for the perovskite top cell. Replacing the (n/p+) junction at the Si wafer's rear with an n/n+ contact suppresses the signal from the bottom silicon cell while keeping the exact same processing steps and overall structure for the perovskite top cell. This assures device-relevant properties of the perovskite cell stack. The results are plotted Figure 2B/D. First, at lower energies (Figure 2B), we confirm that the SPV signal emerging from the silicon bottom cell was effectively suppressed. Similarly, to the early rise for the SHJ (Figure 2A), we do observe slight negative signal for 1.4 and 1.5 eV excitations. Again, we ascribe this to the elimination of the weaker front n/n+ band bending, accompanied by injection of electrons from the n-silicon into the n+ contact. It is of interest that for the longer wavelengths, no such signal is observed. There the small front and rear opposing built-in fields are both suppressed, leading to SPV signals that cancel out, which we ascribe to be due to a more homogeneous generation profile.

Finally, in Figure 2D, at higher excitation energies (Si + Pero excitation), we show the presence of a negative SPV signal. In comparison to Figure 2B, we do not observe any delayed SPV rise at 1μs. Therefore, the observed signal mainly originates from the top perovskite solar cell, with much reduced contribution from the bottom silicon cell. We also confirm this by showing that the tr-SPV signal of a perovskite/2PACz sample deposited on ITO glass (Figure S4) is comparable to the "isolated top cell" signal of Figure 2D. There, we observe a reduction of the plateau at early times for the glass/ITO/HTL/perovskite as compared to the symmetric SHJ/ITO/HTL/perovskite sample. We relate this loss of SPV to the possible re-injection of extracted holes into the perovskite layer when there are no extra free electrons available in the ITO deposited on glass. In contrast, the recombination ITO of the tandem stack should contain extra free electrons from the excited back silicon cell, inducing electron (silicon extracted) hole (perovskite extracted) recombination and resulting in a more sustained SPV. However, other mechanisms are also possible to explain the differences of Figure S4 and exact identification is out of the scope of this work.

To conclude, we demonstrate that with wavelength-dependent SPV measurements, one can isolate the contributions of the perovskite front and silicon bottom cells in such tandem architectures. Additionally, we show that the use of a substrate with symmetric contacts – in our case, an n+/n/n+ silicon heterojunction - effectively suppresses the signal of bottom silicon solar cells while preserving the processing sequence as well as SHJ top contact serving as the perovskite substrate, and thus the properties of the perovskite top cell. trSPV on such a symmetric substrate thus results in the probing of a highly isolated and device-relevant perovskite top cell, paving way for effective optimization of the top perovskite cell stack as well as process parameters.

### 2.2 Intensity-dependent SPV

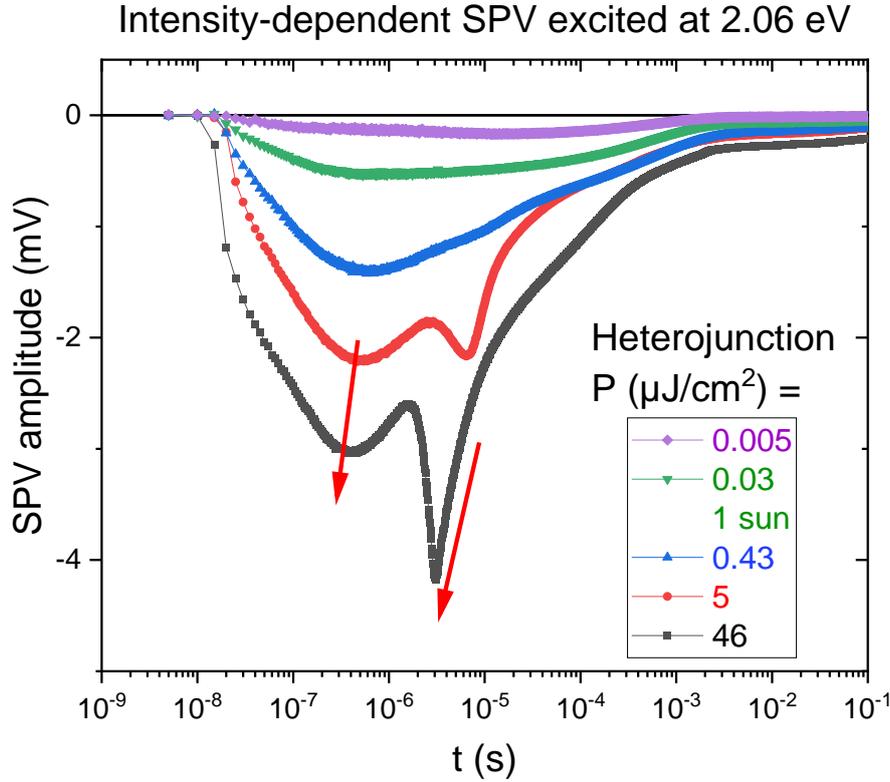

Figure 3: Intensity-dependent transient SPV measurements for perovskite on a silicon heterojunction cell with front side illumination at 620 nm laser excitation at 2 Hz repetition rate.

To better understand the interplay between the charge carrier generation in both absorbers, and its effect on the band bending, in Figure 3, the intensity was varied at a constant 620 nm (2 eV) excitation wavelength to excite both absorbers simultaneously (according to the EQE plots shown in Figure S2, 87.5% absorption by the perovskite, and 3.0% by the silicon). The equivalent photogenerated carrier concentrations range from $5x10^{18}$ cm$^{-3}$ for 46 µJ/cm² to $5x10^{14}$ for 0.005 µJ/cm². We note that the equivalent laser fluence for 1-sun illumination is 0.03 µJ/cm² (green curve), corresponding to $3x10^{15}$ cm$^{-3}$. From 0.005 to 0.43 µJ/cm², the SPV recombination dynamics were similar to that of a single-junction perovskite cell [10], suggesting that nearly all photogeneration occurs within the perovskite layer.

The transients shown in Figure 3 exhibit a combination of charge extraction processes from the perovskite absorber and a shifted delayed SPV peak originating from the Si bottom cell. At an intensity of 0.43 µJ/cm², the major contribution to the SPV signal comes from the perovskite layer, as can be seen from the peak time of 0.59 µs in agreement with the extraction times observed for single-junction perovskite-only half cells utilizing 2PACz as an HTL.[10] At higher intensities, from 5 to 46 µJ/cm², we observe the appearance of an ambipolar diffusion peak. The peak onset decreases from 7.8 to 3.7 µs. This reduction in peak onset time is explained by inspecting the electron and hole arrival times at the back of the Silicon wafer. Using our diffusion model (note

S8 in the SI), we observe that a reduction of charge arrival time is expected as the incoming photon flux increases as shown in Figure S7, corroborating the experimental observations of Figure 3. Also, we show that the reduction of arrival time is solely due to the diffusion term of ES1 and ES2 (see Figure S8): therefore, the reduction of peak SPV time is a signature of the ambipolar diffusion of carriers within the c-Si wafer.

## 3   Charge separation model deduced from SPV transients over a wide time range

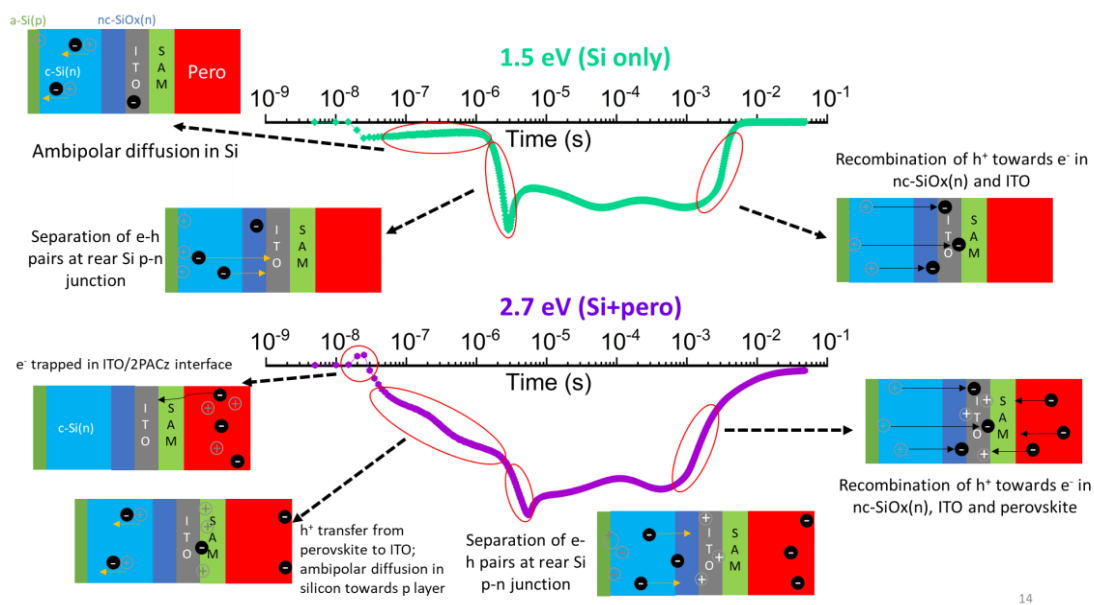

Figure 4: Summary of the different suggested phenomena explaining the SPV transients over a wide time range for a perovskite / n-type Heterojunction tandem layer stack for 1.5 eV (Si-only excitation, green curve, top), and 2.7 eV excitation (Si+pero excitation, purple curve, bottom).

Finally, Figure 4 illustrates the processes that we have considered in our transients. The case of low energy photons ($E_{phot} < 1.3eV$) is not depicted here, as it does not contain any valuable information on the perovskite cell: The perovskite is transparent at this photon energy, and photogenerated charges are close to homogeneously distributed in the Si absorber. This leads to an immediate reduction of band bending at the rear p+/n contact, thus a sharp increase in SPV, followed by a monotonous decrease in the signal. This behavior is fully governed by the c-Si bottom cell's properties. Next, we consider a photon energy of 1.5eV (Figure 4, upper panel). Here, the perovskite is still transparent, whereas the absorption depth, $\alpha^{-1}$, in c-Si is ca. 40 μm. The initial negative SPV signal at ~ 20 ns originates from photogenerated charge carriers that undergo separation close to the recombination ITO layer. Because the a-Si:H (p) layer is located 250 μm away at the rear, ambipolar diffusion of the remaining electron-hole pairs to the rear side occurs on the timescale of ~2-4 μs. When the electron-hole pairs reach the rear side selective layer (hole contact), the electron-hole pairs are immediately separated at the rear junction, giving rise to the second, delayed negative SPV signal at ~ 2 μs. The following slow decay corresponds to the recombination within the Si bottom cell and/or due to the ambipolar diffusion of carriers (where slower photogenerated holes redistribute after electrons, effectively "catching up" electrons). For the 2.7 eV excitation (Figure 4, bottom purple curve), a small positive signal is observed at around 10 ns, which could be explained by electron trapping at the ITO/2PACz interface.[10] In the regime between 30 ns–1 µs, the signal is composed of charge separation by the perovskite SAM interface and by the separation of electrons photogenerated in the Si towards the recombination ITO. During the first few µs, ambipolar diffusion occurs within the c-Si layer, again giving rise to the second delayed negative SPV peak at ca. 3-6 µs, followed by delayed charge separation by the rear side in the junction with the a-Si(p). The slow decay after 6 µs corresponds to back transfer and recombination of the remaining charge carriers in the different layers.

Due to the complexity of the cell architectures studied here, we refrain from establishing a direct correlation between the tr-SPV kinetics and the actual tandem solar cell PV parameters   (shown in Figure S5). However, future studies will be focused on addressing the influence of the individual layers on the tr-SPV kinetics and solar cell parameters simultaneously.

## 4. Conclusion

In this work, perovskite films with a band gap of 1.68 eV slot-die coated on heterojunction and symmetric heterojunction Si bottom cells were investigated by wavelength-dependent time-resolved surface photovoltage measurements. The results showed delayed charge separation for lower energy photons in the silicon bottom cell. Ambipolar diffusion of electron-hole pairs to the rear p-n junction of the heterojunction Si bottom cell was found to be the cause of a delayed SPV signal. At high photon energies, a symmetric SHJ tandem showed an isolated signal from the top perovskite cell whereas the SHJ tandem showed a convoluted signal from the bottom and top cells. Two main conclusions are drawn towards the results. First, with photon energies lower than the band gap of the perovskite top cell, we successfully isolate the bottom silicon cell. Second, using a symmetric SHJ bottom silicon cell, we demonstrate an isolated SPV signal originating from the top perovskite cell. Thus, an in-depth analysis of the charge carrier dynamics of a tandem-equivalent perovskite cell stack is demonstrated, using spectrally resolved tr-SPV with the combination of symmetric and non-symmetric bottom c-Si cells.

With the ability to effectively isolate single interfaces from the bottom or the top cell, the methodology shown here can be applied to gain valuable insights into the carrier dynamics of different 2T tandem architectures to pin-point losses at interfaces and select appropriate charge transport layers for the perovskite top cell, in-route for tandems approaching their theoretical limit.

**Acknowledgements**
We thank Thomas Dittrich for fruitful discussions and for providing SPV lab facilities for the tr-SPV measurements, and Isaac Balberg and Hannes Hempel for fruitful discussions. The authors acknowledge the Helmholtz Association for funding within the HySPRINT Innovation lab and the Zeitenwende project as well as the Federal Ministry of Education and Research (BMWK) for funding the projects SHAPE (grant no. 03EE1123C), as well as funding from the German State of Lower Saxony in project "NextGenPV", through funding program "zukunft.niedersachsen". M.S. is grateful for support by the German-Israeli Helmholtz International Research School HI-SCORE (HIRS-0008).

**Author contributions**
M.S. and K.X. contributed equally to this work. K.X. and I.L. initiated the research and planned the experiments. K.X. prepared the perovskite top cell on different Si substrates; K.X., I.L. and M.S. performed the tr-SPV-measurements. K.X., I.L. and M.S. analyzed all the data and took the lead in drafting the manuscript. L.K advised on the interpretation of results and on modelling. M.S. performed the simulations. K.X. and M.S. wrote the paper from the input from other authors. All authors contributed to the discussion of the results.

**Declaration of interests**
The authors declare no competing interests.

# Monitoring charge separation of individual cells in perovskite/silicon tandems via transient surface photovoltage spectroscopy

Maxim Simmonds [1], Ke Xu [1], Steve Albrecht [1,2], Lars Korte*[1,2], Igal Levine*[3]

[1] Solar Energy Division, Helmholtz-Zentrum Berlin für Materialien und Energie GmbH, Kekulestraße 5, 12489 Berlin, Germany
[2] Faculty of Electrical Engineering and Computer Science, Technical University Berlin, Marchstraße 23, 10587 Berlin, Germany
[3] Institute of Chemistry and The Center for Nanoscience and Nanotechnology, The Hebrew University of Jerusalem, Jerusalem 9190401, Israel

*Email: igal.levine@mail.huji.ac.il, korte@helmholtz-berlin.de

**Tr-SPV measurements:**

Transient SPV spectroscopy measurements were performed using an oscilloscope card (Gage, CSE 1622-4GS) and a tunable Nd:YAG laser for excitation (duration time of laser pulses 3–5 ns, range of wavelengths 216–2600 nm (EKSPLA, NT230–50, equipped with a spectral cleaning unit). The repetition rate of the laser pulses was 2 Hz, and 20 transients were averaged. The SPV measurements were performed in the configuration of a parallel plate capacitor (quartz cylinder partially coated with the $SnO_2$:F electrode, mica sheet as an insulator) with a high-impedance buffer. For intensity-dependant measurements, the laser fluence was controlled using ND filters. For wavelength-dependant measurements, a variable beam expander was used to keep a relatively constant photon wavelength, as shown in Figure S1.

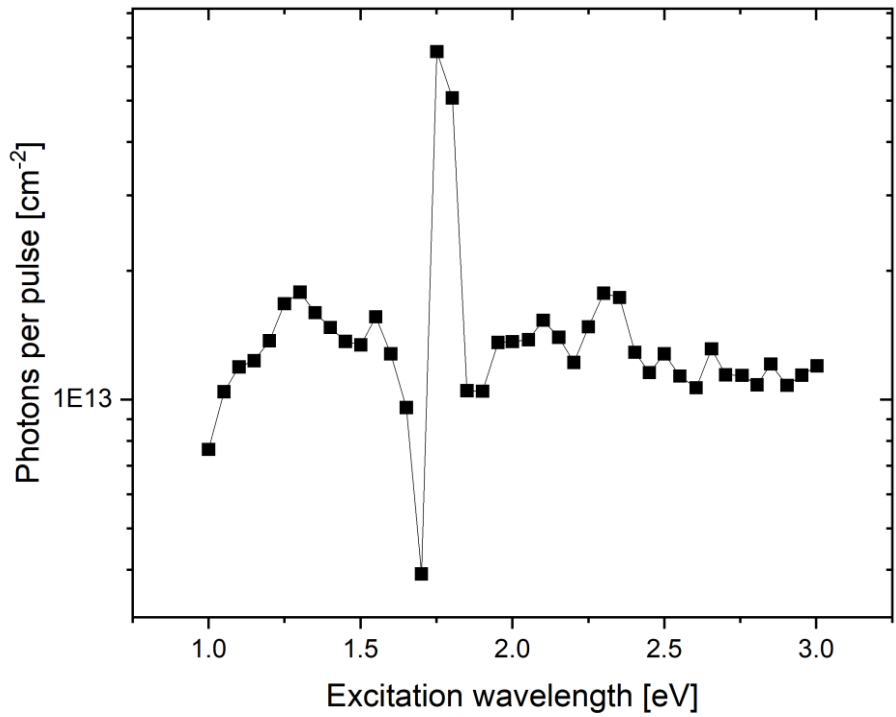

Figure S1: The laser fluence as a function of photon energy.

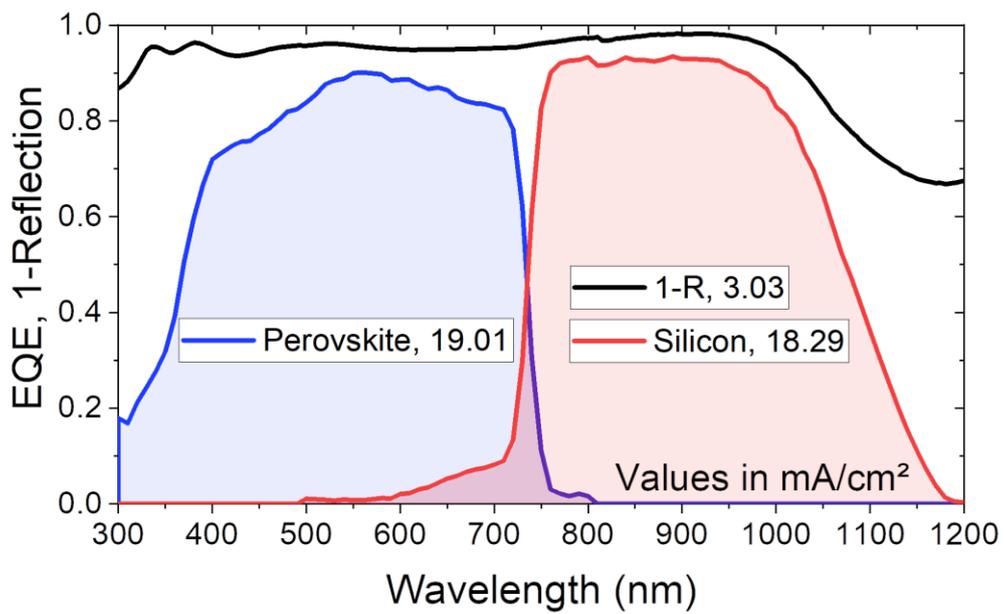

Figure S2: The tandem solar cell's EQE and UV-Vis (1-R, black line) measurements for SHJ with slot-die coated 3halide perovskite. The perovskite absorption rises from 800 nm (1.6 eV). When the wavelength is above 800 nm, all the light is purely absorbed by the silicon solar cell.

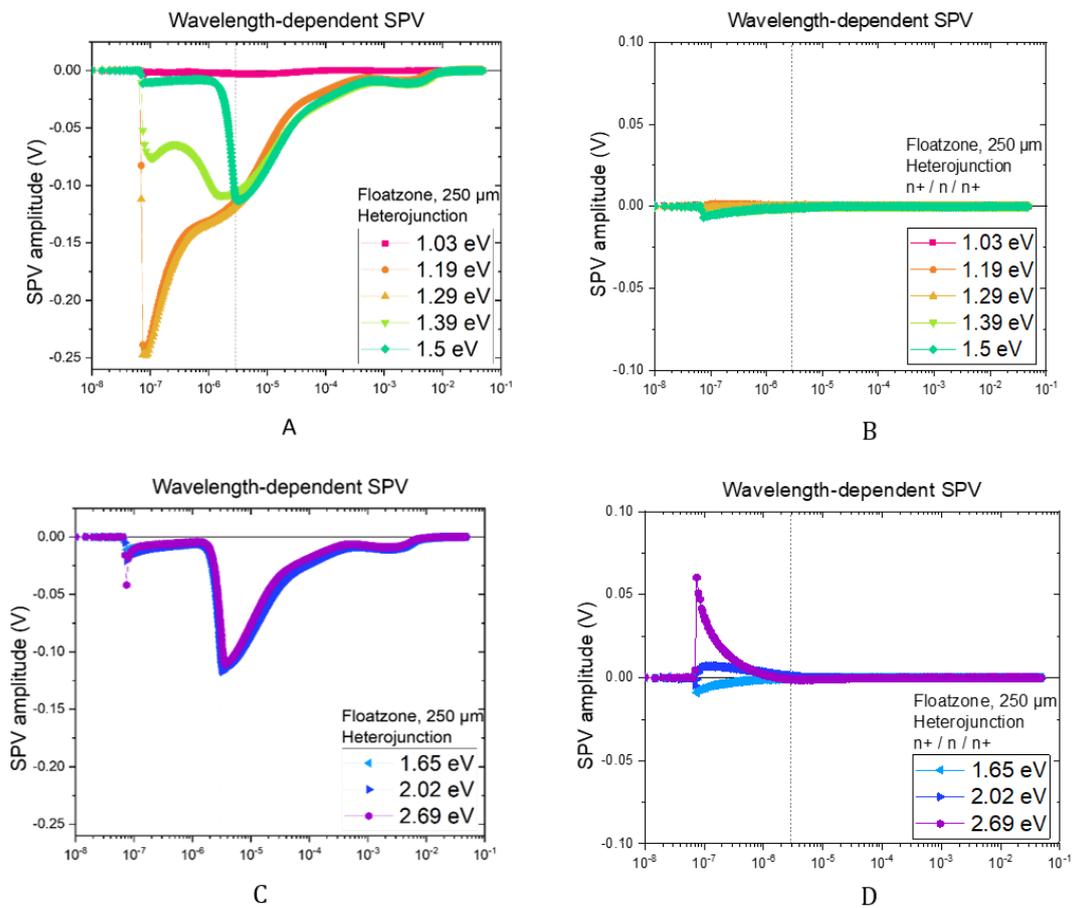

Figure S3: Time-resolved, wavelength-dependent SPV spectra as a function of the excitation wavelength for the SHJ without perovskite layer on top from 1-1.5 eV (A) and 1.65-2.7 eV (C) and for the symmetric SHJ (without perovskite layer on top) from 1-1.5 eV (B) and 1.65-2.7 eV (D).

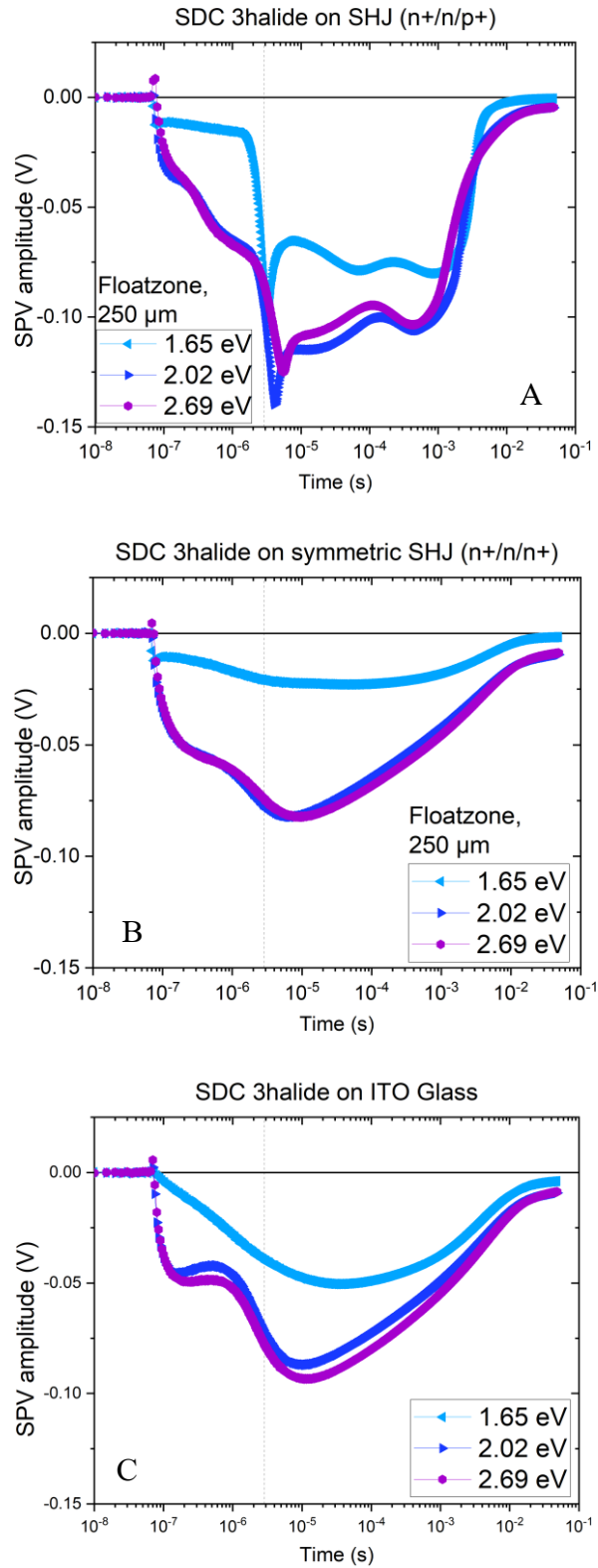

Figure S4: Time-resolved, wavelength-dependent SPV spectra as a function of the excitation wavelength for (A) SHJ/ITO/HTL/perovskite, (B) symmetric-SHJ/ITO/HTL/perovskite and (C) glass/ITO/HTL/perovskite with front side illumination at 2 Hz repetition rate. As shown, the "parasitic" signal coming from absorbed photons in the silicon substrate has been highly reduced (A vs B).

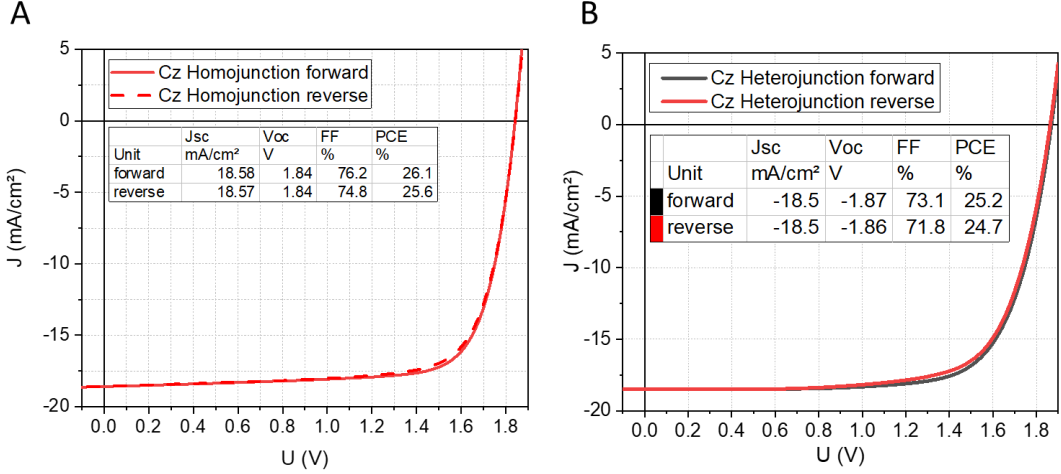

Figure S5: (A) J-V curves of the champion slot-die coated perovskite/silicon tandem solar cell on homojunction silicon bottom cell. Red dash line: reverse scan (from *Voc* to *Jsc*); red solid line: forward scan. (B) J-V curves of the champion slot-die coated perovskite/silicon tandem solar cell on heterojunction silicon bottom cell. Red line: reverse scan (from *Voc* to *Jsc*); black line: forward scan.

**Supplementary Note S6: Drift simulations**

We use a set of differential equations to describe the ambipolar diffusion of carriers within a crystalline silicon wafer after photoexcitation, including the auger and S.R.H non-radiative recombination terms:

$$\frac{dn(z)}{dt} = -k_{rad}\left(n(z)p(z) - n_i^2\right) - \frac{n(z)p(z) - n_i^2}{n\tau_n + p\tau_p} - \left(C_n n + C_p p\right)\left(n(z)p(z) - n_i^2\right) +$$

$$D_n \frac{\delta^2 n(z)}{\delta z^2} \text{(ES1)}$$

$$\frac{dp(z)}{dt} = -k_{rad}\left(n(z)p(z) - n_i^2\right) - \frac{n(z)p(z) - n_i^2}{n\tau_n + p\tau_p} - \left(C_n n + C_p p\right)\left(n(z)p(z) - n_i^2\right) +$$

$$D_p \frac{\delta^2 p(z)}{\delta z^2} \text{(ES2)}$$

Where $n(z)$ ,$p(z)$ $[m^{-3}]$ are the carrier densities in function of the depth of the sample and time. $n_i$ $[m^{-3}]$ is the intrinsic carrier concentration. $k_{rad}$ $[m^3 s^{-1}]$ is the radiative recombination term. $\tau_n$ and $\tau_p$ $[s^{-1}]$ are the hole and electron lifetimes. $C_n$ and $C_p$ $[m^6 s^{-1}]$ are the electron and hole auger recombination coefficients. $D_a$ $[m^2 s^{-1}]$ is the ambipolar diffusion coefficient. With this model, we assume that at the high photon fluxes and thus carrier densities in the sample, an Auger recombination term is needed. For the non-radiative S.R.H., we use the full description.

To calculate the initial conditions for the carrier densities $n_0(z), p_0(z)$ we use the lambert-beer relation (ES2) and multiply by the pulse length of our laser:

$$G(z) = \alpha N_0 e^{-\alpha z} \quad \text{(ES3)}$$
$$n_0(z) = p_w G(z) + N_d \quad \text{(ES4)}$$
$$p_0(z) = p_w G(z) \quad \text{(ES5)}$$

Where $\alpha\,[m^{-1}], N_0[m^{-2}s^{-1}]$ are the absorption coefficient and photon flux. $p_w\,[s]$ is the pulse width and $N_d[m^{-3}]$ is the donor dopant density. The doping density is considered by adding $N_d$ in ES4. The parameters used for simulations are tabulated in Table S1.

| Parameter Name | Symbol | Value | Unit | Source |
|---|---|---|---|---|
| Temperature | $T$ | 300 | $[K]$ | - |
| Intrinsic carrier density | $n_i$ | $10^{10}$ | $[cm^{-3}]$ | [1] |
| Bimolecular radiative recombination coefficient | $k_{rad}$ | $5\ 10^{-15}$ | $[cm^3 s^{-1}]$ | [2] |
| Electron, Hole S.R.H recombination lifetimes | $\tau_n\,,\tau_p$ | $8\ 10^{-3}, 8\ 10^{-3}$ | $[s]$ | Estimation from in-house measurements |
| Electron, Hole Auger recombination coefficients | $C_n, C_p$ | $2.3\ 10^{-31}, 9.9\ 10^{-32}$ | $[cm^6 s^{-1}]$ | [3] |
| Electron, Hole and ambipolar Diffusion coefficients | $D_n, D_p, D_a$ | $34.5, 11.8, 24.3$ | $[cm^2 s^{-1}]$ | [1], [4] |
| Donor doping density (n doping) | $N_d$ | $1.5\ 10^{15}$ | $[cm^{-3}]$ | Conversion from 3 $\Omega\,cm$ resistivity with [1] |

| Absorption coefficient | $\alpha$ | spectrum | $[cm^{-1}]$ | [5] |
|---|---|---|---|---|
| Photon flux | $N_0$ | 3 $10^{21}$ | $[cm^{-2}s^{-1}]$ | Estimated from setup parameters |
| Pulse width | $p_w$ | 5 $10^{-9}$ | $[s]$ | From setup measurements. |
| Wafer thickness | $L$ | 250 | $[\mu m]$ | Manufacturer |

Table S1: Parameters used for the drift simulation.

**Supplementary Note S7: Simulation results, diffusion in function of photon energy**

To understand the second onset time of SPV due to holes diffusing in the layer and attaining the back contact of the Silicon heterojunction solar cell, we define the threshold arrival time $t_{thresh}$ as the time that the concentration at the back of the wafer (position x = 248 µm) attains 95% of its maximal value. The time evolutions of the carrier densities (2µm from the back) are plotted in Figure S6: As shown, the arrival times are dependent on the power used and the photon energy. In Figure S6C, we see that $t_{thresh}$ increases until 18µs, with a plateau reached at 2 eV. This plateau is within the onset of SPV observed in Fig 2A of the main text (6 µs).

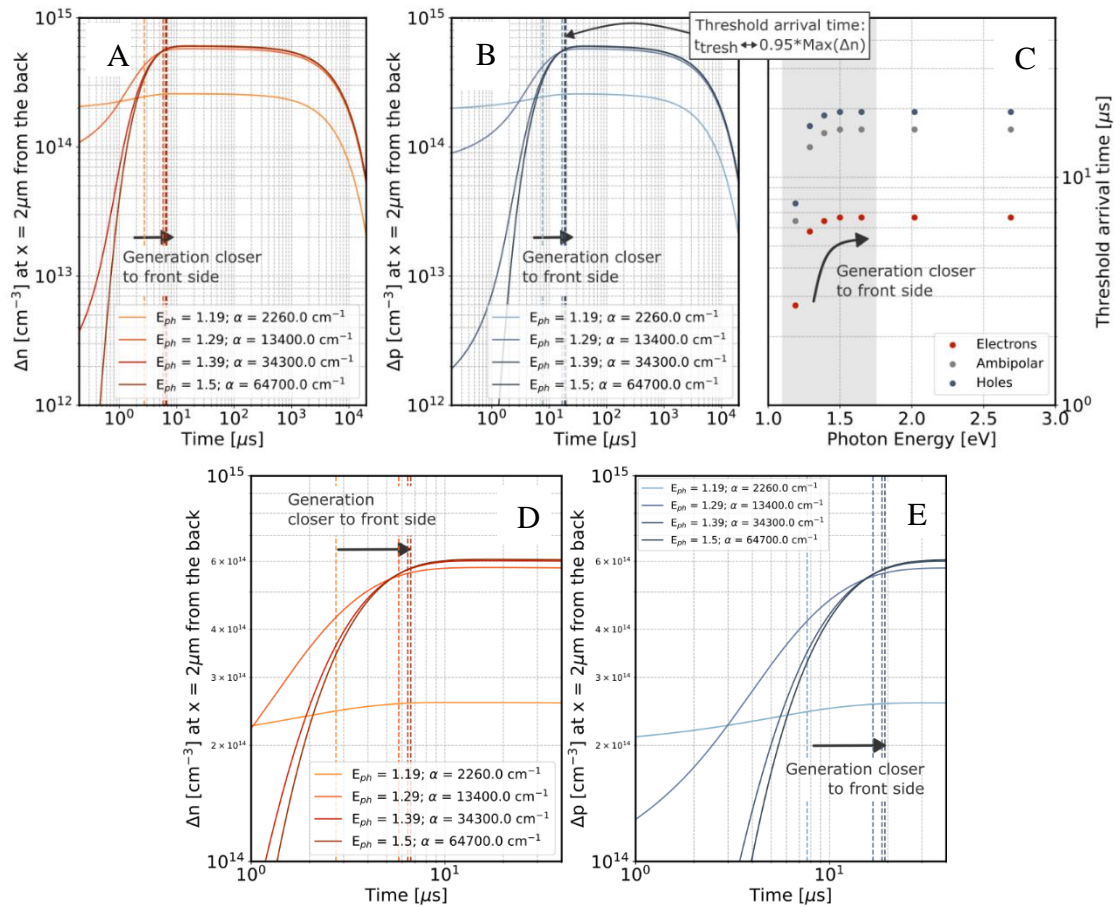

Figure S6: Time trace of photogenerated (A) Electron and (B) Hole carrier density at position x = 248um. (C) Electron, Hole and Ambipolar threshold arrival time in function laser photon energy. We see an increase in the arrival time until saturation at $E_{photon} > 1.7$ eV. The increase is in accordance with Figure 2 of the main text, at photon energies < 2eV. (D) and (E) are scaled-in versions of (A) and (B), respectively.

**Supplementary Note S8: Simulation results, diffusion as a function of incoming photon flux.**

In Figure S7C, we see a power dependence of the threshold arrival time for 2eV photons: there is a reduction of the threshold arrival time when increasing power, consistent with our experimental results shown in Figure 3. Figure S8 shows the threshold arrival time when removing different components of equations ES1 and ES2 in the model. Four cases arise with a combination of the inclusion or not of Auger as well as SRH non radiative recombination, respectively. With this analysis, we conclude that it is the diffusion term that explains the reduction of arrival time: The reduction of arrival time can therefore be interpreted as a signature for ambipolar diffusion.

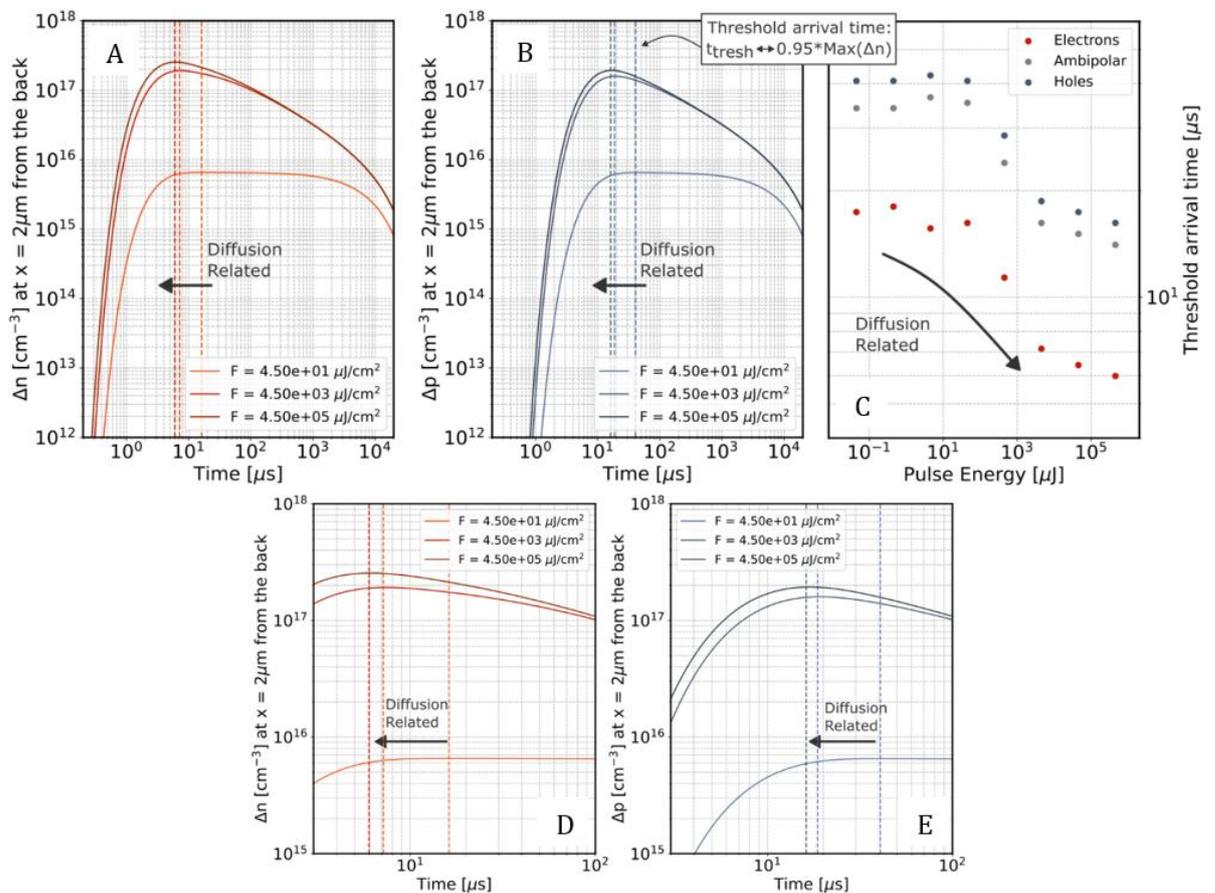

Figure S7: Time trace of photogenerated (A) Electron and (B) Hole carrier density at position x = 248um. (C) Electron, Hole and Ambipolar threshold arrival time in function of pulse energy. We see that the arrival times reduce with pulse energy, in accordance with the observations of figure 3 of the main text. (D) and (E) are scaled-in versions of (A) and (B), respectively.

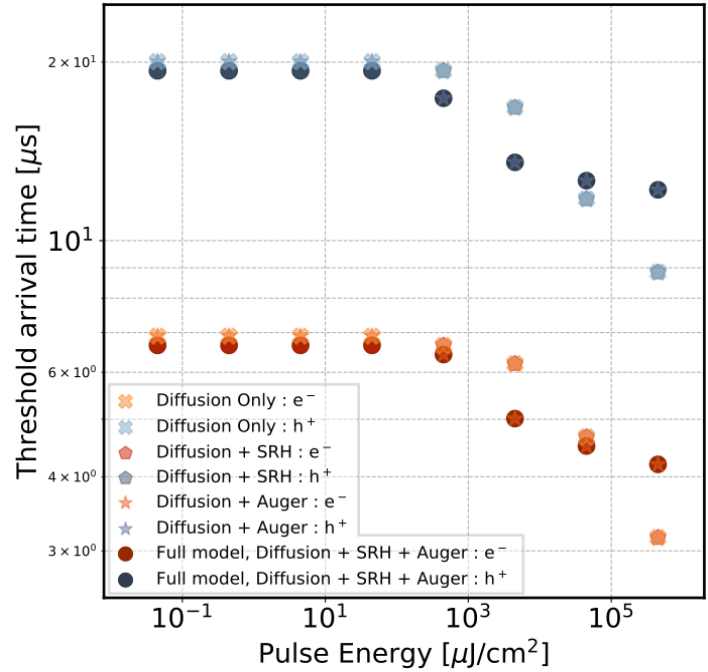

Figure S8: Threshold arrival times by using a mixture of the components of equations ES1 and ES2. We see that the power dependence of the threshold arrival time is mainly due to the diffusion of carriers.